\documentclass[aps,prb,reprint,amsmath,a4paper,superscriptaddress]{revtex4-1}

\usepackage{graphicx}
\usepackage{color}
\usepackage[normalem]{ulem}

\begin{document}

\title{Inter--Edge Backscattering in Buried Split--Gate--Defined \\ Graphene Quantum Point Contacts}

\author{Shaohua Xiang}
\affiliation{NEST, Istituto Nanoscienze--CNR and Scuola Normale Superiore, Piazza San Silvestro 12, 56127 Pisa, Italy}

\author{Alina Mre\'nca--Kolasi\'nska}
\affiliation{AGH University of Science and Technology, Faculty of Physics and Applied Computer Science, al.~Mickiewicza 30, 30-059 Krak\'ow, Poland}

\author{Vaidotas Miseikis}
\affiliation{Center for Nanotechnology Innovation @NEST, Istituto Italiano di Tecnologia, Piazza San Silvestro 12, 56127 Pisa, Italy}

\author{Stefano Guiducci}
\affiliation{NEST, Istituto Nanoscienze--CNR and Scuola Normale Superiore, Piazza San Silvestro 12, 56127 Pisa, Italy}

\author{Krzysztof Kolasi\'nski}
\affiliation{AGH University of Science and Technology, Faculty of Physics and Applied Computer Science, al.~Mickiewicza 30, 30-059 Krak\'ow, Poland}

\author{Camilla Coletti}
\affiliation{Center for Nanotechnology Innovation @NEST, Istituto Italiano di Tecnologia, Piazza San Silvestro 12, 56127 Pisa, Italy}

\author{Bart\l{}omiej Szafran}
\affiliation{AGH University of Science and Technology, Faculty of Physics and Applied Computer Science, al.~Mickiewicza 30, 30-059 Krak\'ow, Poland}

\author{Fabio Beltram}
\affiliation{NEST, Istituto Nanoscienze--CNR and Scuola Normale Superiore, Piazza San Silvestro 12, 56127 Pisa, Italy}

\author{Stefano Roddaro}
\affiliation{NEST, Istituto Nanoscienze--CNR and Scuola Normale Superiore, Piazza San Silvestro 12, 56127 Pisa, Italy}

\author{Stefan Heun}
\email[]{stefan.heun@nano.cnr.it}
\affiliation{NEST, Istituto Nanoscienze--CNR and Scuola Normale Superiore, Piazza San Silvestro 12, 56127 Pisa, Italy}

\date{\today}

\begin{abstract}
Quantum Hall effects offer a formidable playground for the investigation of quantum transport phenomena. Edge modes can be deflected, branched, and mixed by designing a suitable potential landscape in a two--dimensional conducting system subject to a strong magnetic field. In the present work, we demonstrate a buried split--gate architecture and use it to control electron conduction in large--scale single--crystal monolayer graphene grown by chemical vapor deposition. The control of the edge trajectories is demonstrated by the observation of various fractional quantum resistances, as a result of a controllable inter--edge scattering. Experimental data are successfully modeled both numerically and within the Landauer-B\"uttiker formalism. Our architecture is particularly promising and unique in view of the investigation of quantum transport via scanning probe microscopy, since graphene constitutes the topmost layer of the device. For this reason, it can be approached and perturbed by a scanning probe down to the limit of mechanical contact.
\end{abstract}


\maketitle

\section{Introduction}

The quantum Hall (QH) effect has offered exciting opportunities for the investigation of quantum transport in two--dimensional electron gas systems\cite{Klitzing1980} for more than three decades, and it still is the foundation for a number of research activities. QH physics is particularly interesting --- for a set of different reasons --- in the case of graphene, \cite{Novoselov2004,Novoselov2005} a two--dimensional (2D) layer of carbon atoms arranged in a honeycomb lattice. First of all, owing to the non--trivial Berry phase of the electron system,\cite{Ando1998,Peres2006} the QH effect in graphene displays half--integer plateaus and thus differs from what is observed in other conventional 2D systems.\cite{Novoselov2004,Novoselov2005,Zhang2005,Zhang2006,Jiang2007,Bolotin2009,Xiang2016} In addition, graphene is an ambipolar material, and opposite QH chiralities can be obtained on the same sample by simply tuning the carrier density: for instance ambipolarity was exploited to investigate Klein tunneling\cite{Katsnelson2006,Gorbachev2008,Stander2009} and the QH physics in graphene $p$--$n$ junctions.\cite{Williams2007,Huard2007,Ozyilmaz2007} Finally, graphene implements a stand--alone one--atom--thick 2D electron system, and charge conduction essentially occurs at its surface. Differently from other materials, conducting electrons can thus be approached down to any small distance: this characteristics offers unique perspectives in view of the investigation of the local conduction properties, in particular in the context of quantum transport and QH physics. In order to take advantage of all these features, however, it is necessary to implement new methods to control the local carrier density in graphene, while retaining direct access to its surface and maintaining low-level disorder. To this end, here we investigate a buried split--gate architecture in which graphene constitutes the topmost layer of the device. We demonstrate that good mobility can be obtained using single--crystal monolayer graphene grown by chemical vapor deposition (CVD).\cite{Miseikis2015,Xiang2016} The successful control of edge trajectories in the QH regime is demonstrated by the observation of fractional quantized resistance values, akin to what was recently demonstrated in a conventional top split--gate architecture (i.e.~a device in which the split--gate is placed on top of the graphene).\cite{Nakaharai2011} Owing to the different electrostatics of the device, though, lever arms values observed here for back and local gates are markedly different with respect to previous works.\cite{Nakaharai2011,Zimmermann}

The observed experimental behavior is described within the framework of the Landauer-B\"uttiker formalism by assuming the presence of three distinct filling factors: in the bulk of the sample ($\nu_{BG}$), above the split--gate electrodes ($\nu_{SG}$), and in the quantum-point-contact opening ($\nu_{QPC}$). In order to support the consistency of this interpretation and to provide a more general modeling framework, we present numerical calculations where the electrostatic potential landscape induced by the electrodes is directly obtained by solving the Poisson equation, and no assumptions are made on local filling factors. In the quantum scattering problem, decoherence and equilibration are introduced in the present model only in regions where different edge modes 
co-propagate.\cite{Ozyilmaz2007, Abanin2007} Numerical results reproduce the experimental data and confirm that transport in our devices is governed by currents flowing along the edges of regions with three distinct filling factors.

\section{Experimental Methods}

\begin{figure}[tb!]
  \includegraphics[width=\columnwidth]{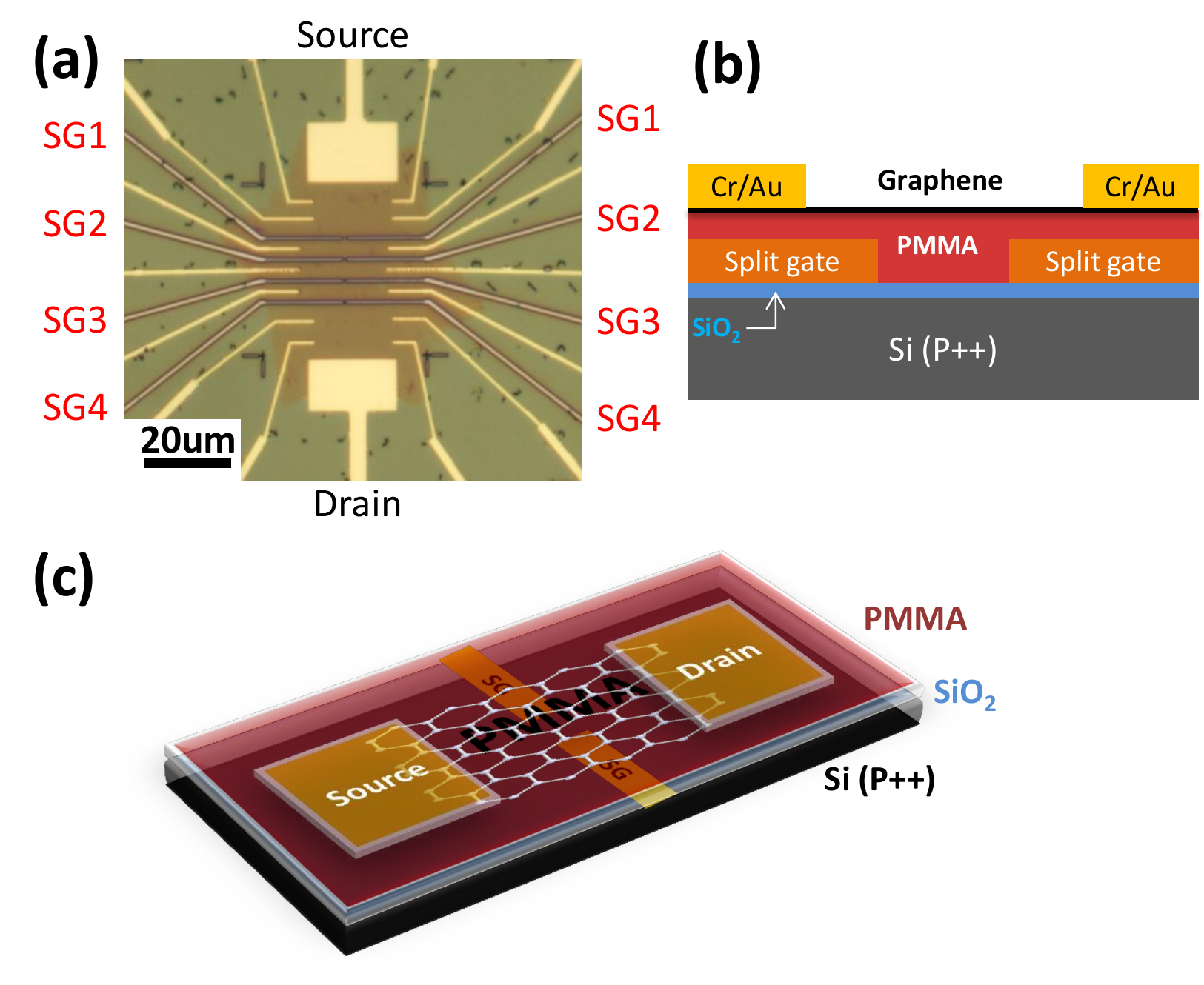}
  \caption{\label{SF1} (a) Optical micrograph of the device. (b) Cross--sectional sketch of the device. (c) Three--dimensional rendering of the layers of the device.}
\end{figure}

In our sample, four split--gates (indicated by red numbers in Fig.~\ref{SF1}(a)) were patterned by electron beam lithography (EBL) on a Si/SiO$_2$ substrate (oxide thickness 300~nm). The width of the gate fingers is 500~nm, while their relative distance is 400, 600, 800, and 1000~nm for each split--gate pair, respectively. The local gating structure was then buried under a PMMA layer, which was spin--coated on the sample and played the role of gate insulator. A 200~$\mu$m $\times$ 200~$\mu$m region centered on the spit--gates was then cross--linked by a high dose e--beam exposure (15000~$\mu$C/cm$^2$), while the rest of the PMMA was dissolved in acetone. The final thickness of the PMMA was measured to be 150 nm.

Single--crystal monolayer CVD graphene was grown on oxidized Cu foil using a cold--wall CVD reactor.\cite{Miseikis2015} To minimize the transfer--induced contamination, it was removed from the growth substrate using electrochemical delamination\cite{Wang2011} and then transferred on top of the PMMA and precisely aligned to the split--gate structures.\cite{Xiang2016} As final step, metallic contacts (Cr/Au: 10~nm/60~nm) to the graphene flake were defined by EBL and thermal evaporation. A cross--section of the complete buried split--gate architecture is shown in Fig.~\ref{SF1}(b). For the present work, only the 400~nm--wide and the 800~nm--wide split--gates were investigated. Measurements were performed using a four--terminal lock--in technique in a $^3$He closed--cycle system with base temperature of 250~mK. The longitudinal and transversal resistances are defined as $R = V_{ij}/I_{SD}$, with $V_{ij}$ the voltage drop measured between contacts $i$ and $j$, and $I_{SD}$ the applied source--drain current (10nA for all measurement). 

\section{Results}

\begin{figure}[tb!]
  \includegraphics[width=\columnwidth]{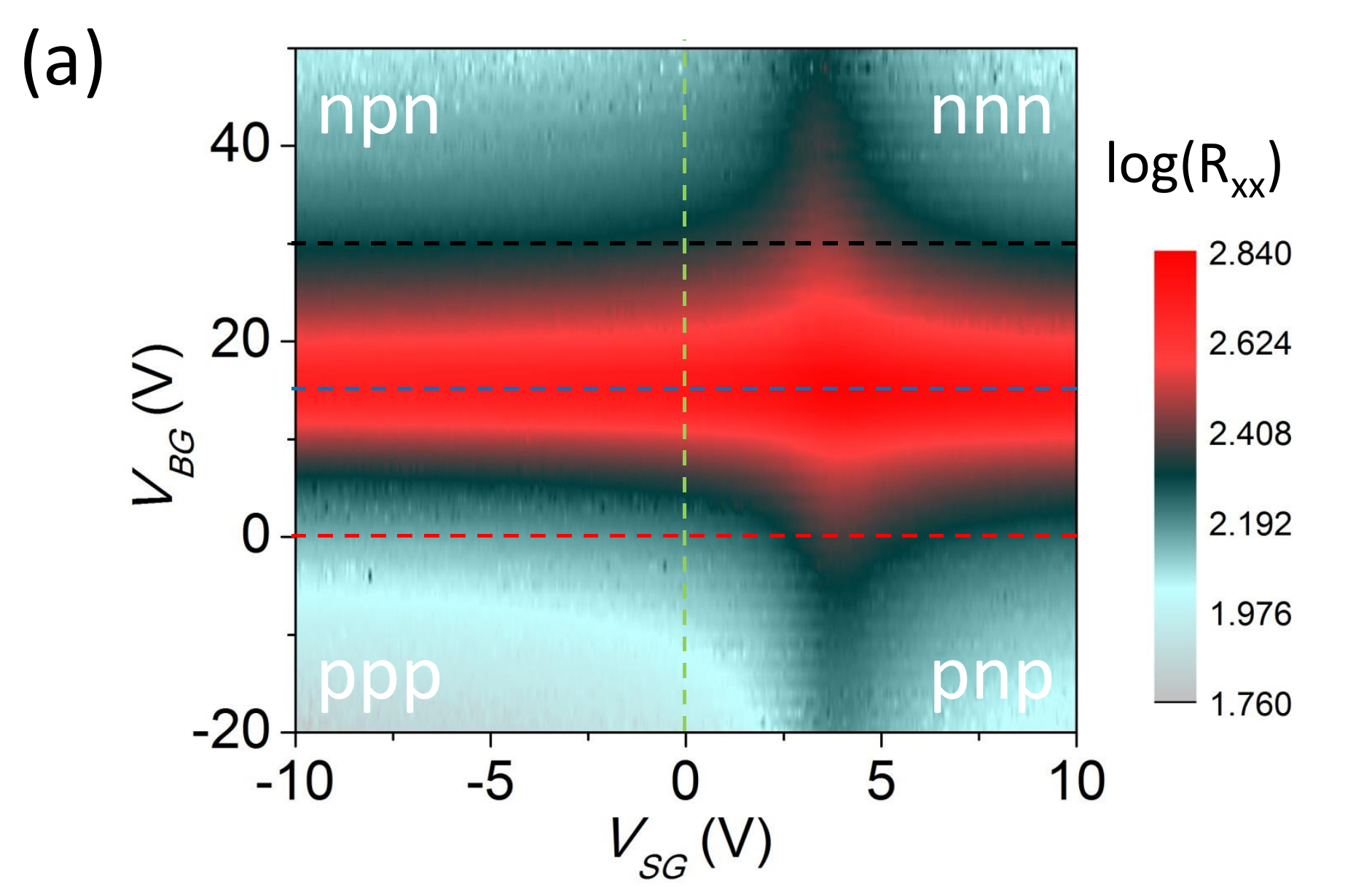}
	\includegraphics[width=0.49\columnwidth]{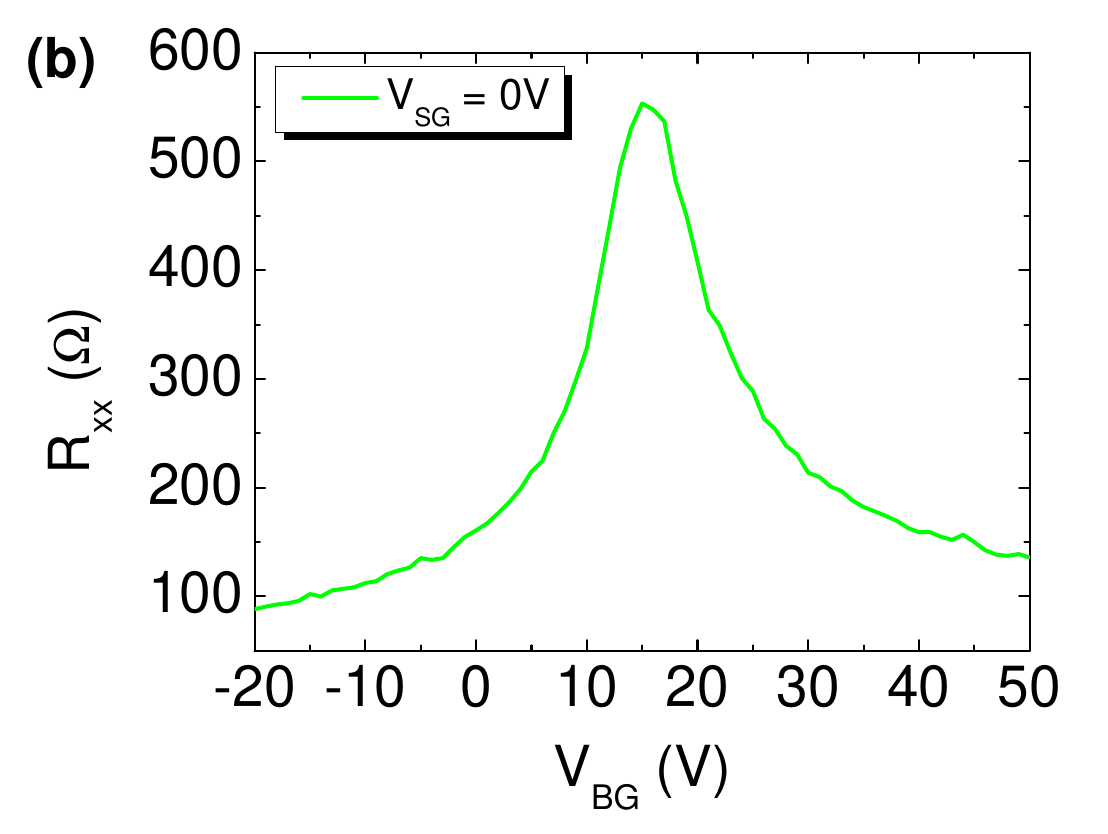}
  \includegraphics[width=0.49\columnwidth]{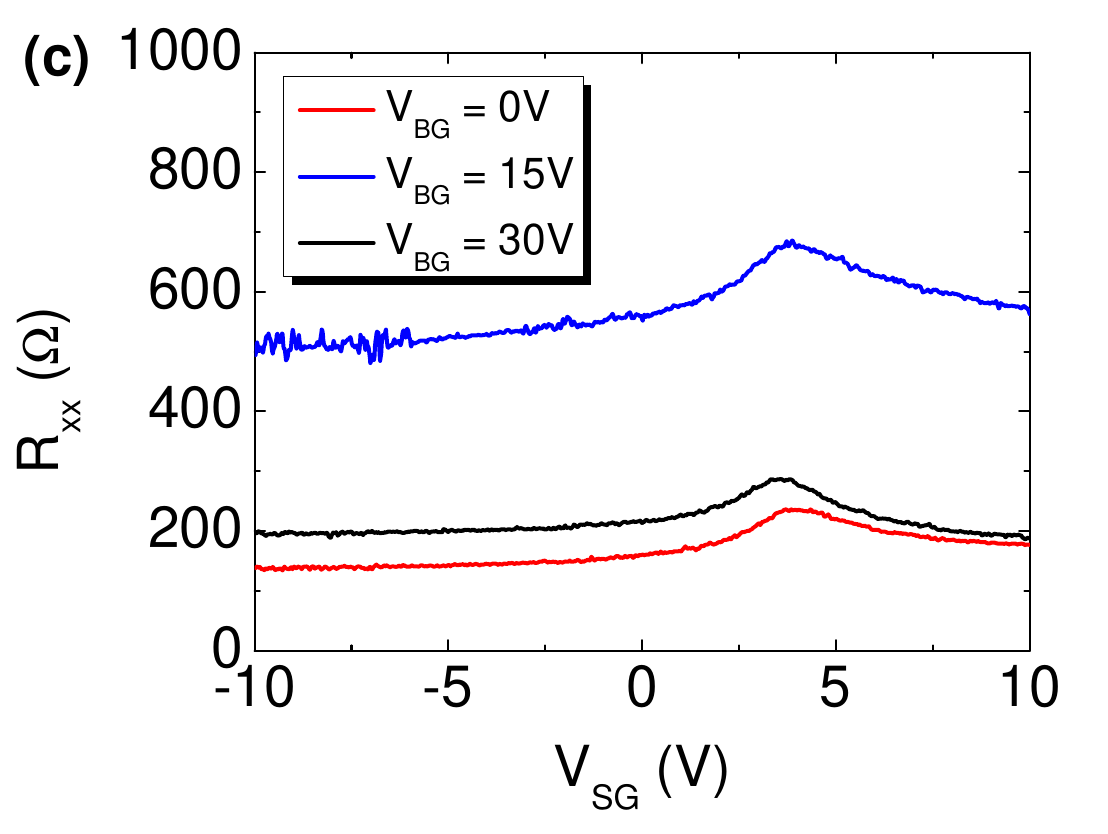}
  \caption{\label{SF2} (a) Measured longitudinal resistance $R_{xx}$ as a function of $V_{BG}$ and $V_{SG}$. Note the logarithmic scale for $R_{xx}$. (b) $R_{xx}$ as a function of $V_{BG}$ for fixed $V_{SG} = 0$~V. (c) $R_{xx}$ as a function of $V_{SG}$ for $V_{BG} = 0$~V, 15~V, and 30~V. Split--gate width 400~nm, B = 0~T, T = 270~mK.}
\end{figure}

The electrostatic action of the various gates on the carrier density in graphene is illustrated in Fig.~\ref{SF2}(a) showing the longitudinal resistance as a function of back--gate ($V_{BG}$) and split--gate ($V_{SG}$) voltages, at B = 0~T. The impact of the back--gate is clearly visible, and a resistance maximum at $V_{BG} \approx 15$~V is observed for every value of $V_{SG}$. This trend is also visible from the cross--sectional plot in Fig.~\ref{SF2}(b), which was obtained along the vertical dashed green line in Fig.~\ref{SF2}(a) at $V_{SG} = 0$~V. This resistance maximum corresponds to the charge neutrality point (CNP), or Dirac point, in the bulk of the graphene flake.

It is well known that the carrier concentration $n$ in graphene varies with $V_{BG}$ as $n \approx C_{BG} \left| V_{BG} - V_{Dirac} \right| / e$.\cite{Wang2013} Here $C_{BG}$ and $e$ are the gate capacitance per area and the elementary charge, respectively. The gate capacitance for a 300--nm SiO$_2$ oxide is $C_{BG0} = 11.5$~nF/cm$^2$.\cite{Wang2013} Here, we need to consider the additional 150~nm thickness of the PMMA layer, which has approximately the same dielectric constant value of SiO$_2$. We therefore set $C_{BG} \approx C_{BG0} / 1.5 = 7.67$~nF/cm$^2$. This yields an intrinsic carrier density of $7.2 \times 10^{11}$~cm$^{-2}$ at $V_{BG} = 0$~V (hole doping), in good agreement with what is typically observed for this kind of CVD graphene.\cite{Xiang2016} The mobility $\mu$ of the device was determined from $\mu = 1 / n e \rho$, where $\rho$ is the resistivity at $B = 0$~T. A transport mobility, away from the Dirac point, of 15300~cm$^2$/(Vs) was determined for $V_{BG} = 0$~V.

The effect of $V_{SG}$ on the carrier density in graphene is not as pronounced, but a peak in $R_{xx}$ can also be observed for every given value of $V_{BG}$. Selected profiles are shown in Fig.~\ref{SF2}(c) and correspond to the horizontal dashed lines in Fig.~\ref{SF2}(a). The resistance maximum is always observed at $V_{SG} \approx 4$~V, regardless of the value of $V_{BG}$. This behavior can be interpreted as due to the local modulation of the carrier density in the regions immediately above the split--gate structure. Also in this case, the resistance peaks when graphene is tuned to the CNP. However, given the limited graphene area controlled by the split--gates, the magnitude of the peak is markedly smaller than the one observed for the back--gate sweep. 

In addition, it should be noted that the two values (4~V and 15~V) are in good agreement with the different expected capacitive couplings between the split--gate and graphene, and between the back--gate and graphene. In the former case, the dielectric insulation is just due to the 150~nm--thick PMMA layer; in the latter, capacitive coupling is mediated by a stack of 300~nm of SiO$_2$ and, again, 150~nm of PMMA.

Differently from what was obtained in recent experiments using a top--gate architecture,\cite{Williams2007,Huard2007,Ozyilmaz2007} the position of the resistance maximum as a function of $V_{SG}$ does not depend on the value of $V_{BG}$. This indicates that back--gate and split--gates are independent, and no cross--talk is observed between them. This is because the back--gate voltage is screened by the buried metallic split--gate, which is inserted in between the back--gate and the graphene layer. To a good approximation, in our devices the bulk carrier density is only controlled by $V_{BG}$, while the density in correspondence to the split--gates is only controlled by $V_{SG}$. On the other hand, in top--gated devices the local carrier density in proximity of the split--gate structure is affected by both back--gate and top--gate voltages. 

\begin{figure}[tb!]
  \includegraphics[width=\columnwidth]{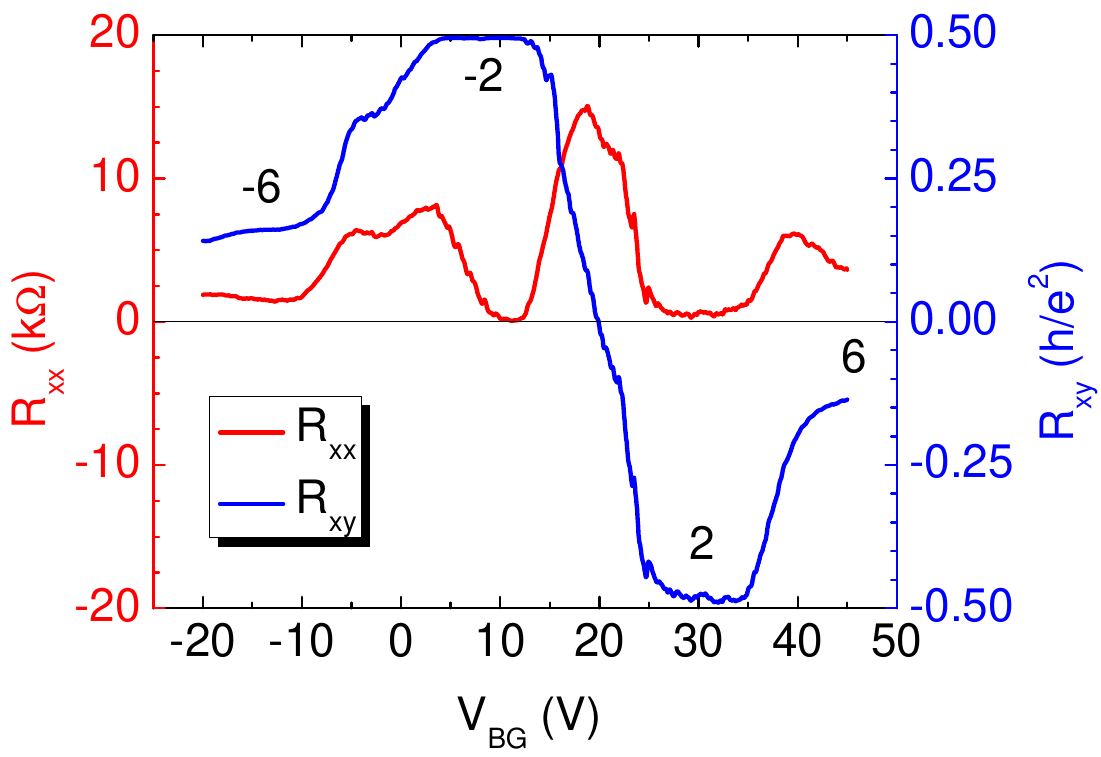}
  \caption{\label{SF3} Longitudinal ($R_{xx}$) and transverse ($R_{xy}$) resistance in the bulk of the graphene flake as a function of back--gate voltage $V_{BG}$ at $B = 10$~T and $V_{SG} = 0$~V.}
\end{figure}

The occurrence of QH states in the bulk of the graphene flake can be inferred from the longitudinal resistance $R_{xx}$ at a magnetic field $B = 10$~T, shown in Fig.~\ref{SF3}, where $V_{BG}$ was swept from $-20$~V to $+45$~V while keeping $V_{SG} = 0$~V. QH plateaus at filling factors $\nu$ = $-6$, $-2$, $+2$, and $+6$ are observed in this $V_{BG}$ range and confirm that the flake is a monolayer and that the graphene quality is good. 

\begin{figure*}[tb!]
  \includegraphics[width=0.48\textwidth]{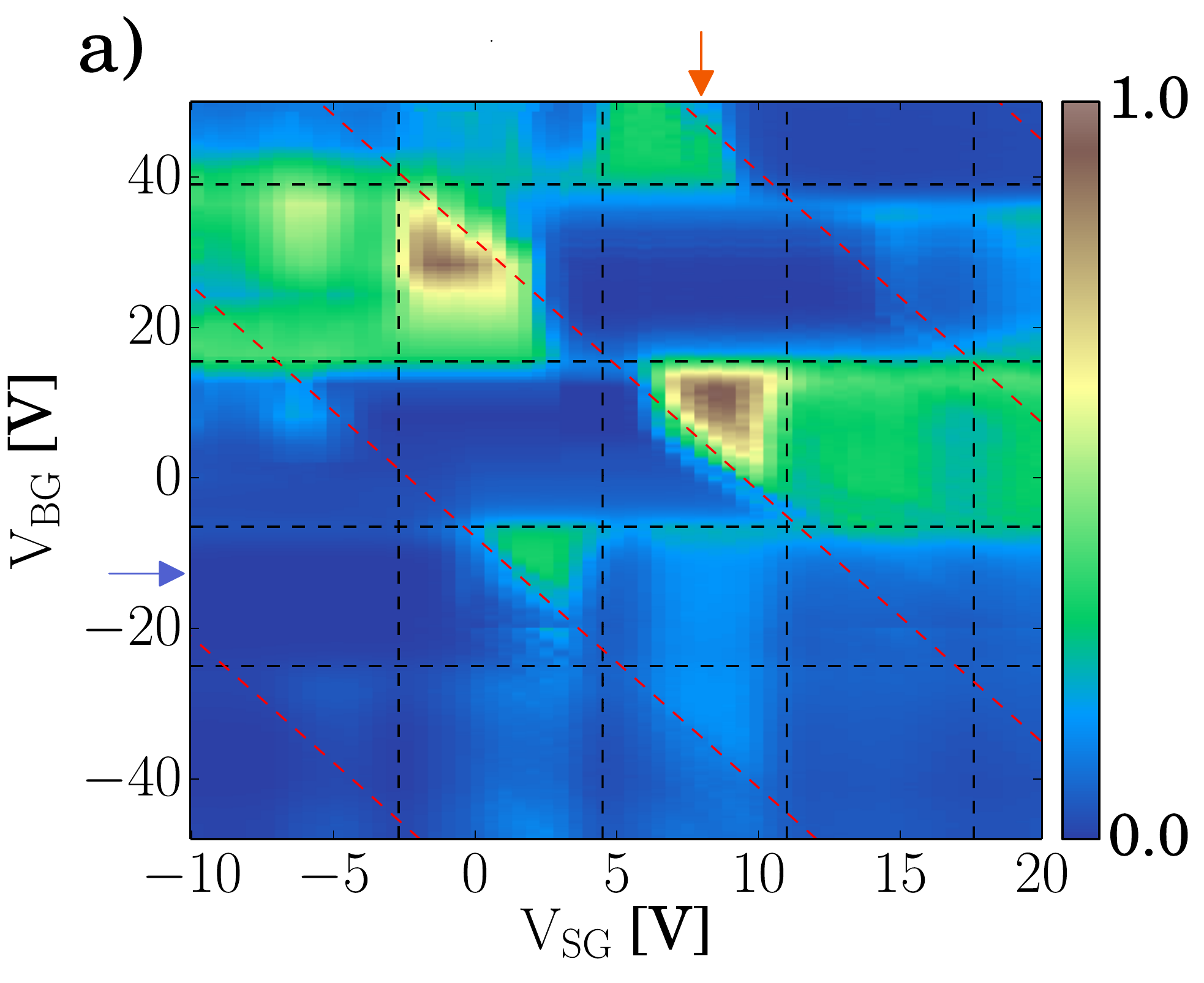}
  \includegraphics[width=0.43\textwidth]{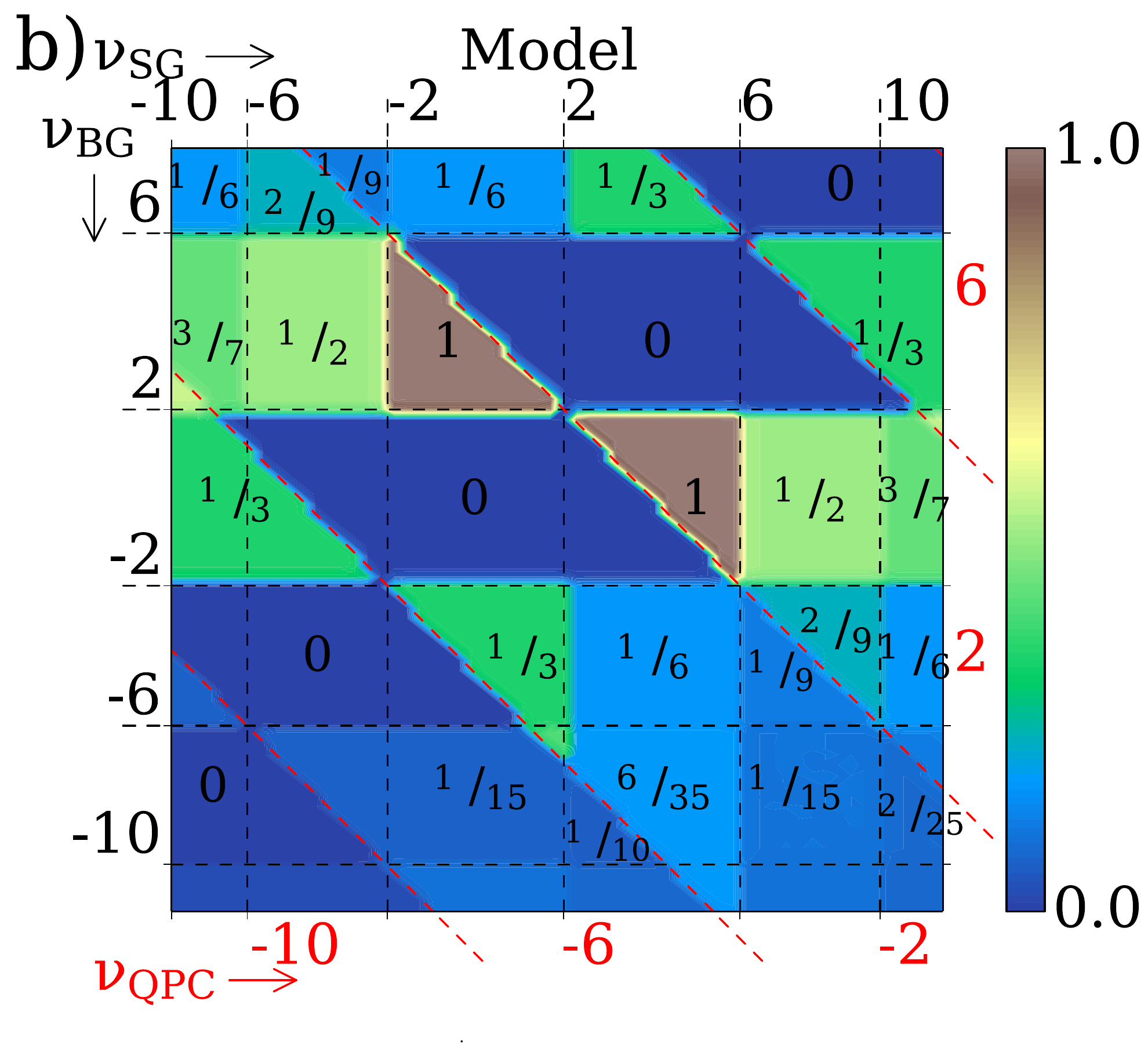}
  \caption{\label{scan} (a) Longitudinal resistance (in units of $h/e^2$) measured across the 400--nm wide QPC as a function of the back--gate ($V_{BG}$) and split--gate ($V_{SG}$) voltage biases at $B = 10$~T. (b) Analytical results obtained with Eqs.~(\ref{eq:unipolar1})--(\ref{eq:bipolar2}) reproduce well the observed resistance plateaus. The arrows in (a) indicate where the cross--sections of Figs.~\ref{cross} (a) and (b) were taken.}
\end{figure*}

The full evolution of the longitudinal resistance $R_{xx}$ across the 400--nm wide QPC as a function of both $V_{SG}$ and $V_{BG}$ is shown in Fig.~\ref{scan}(a): various fractionally quantized  regions can be spotted in the colorplot. As argued in the following, they can be understood in terms of carrier density configurations leading to three different filling factors: the bulk filling factor $\nu_{BG}$ that is controlled by $V_{BG}$, the filling factor in correspondence to the split--gate fingers $\nu_{SG}$ in turn controlled by $V_{SG}$, and the filling factor in the constriction region --- or split--gate opening --- $\nu_{QPC}$ that is driven by both gates.

In contrast to the results obtained with $n$--$p$--$n$ junctions\cite{Ozyilmaz2007,Williams2007,Huard2007,Abanin2007,Ki,Zimmermann} or quantum point contacts (QPCs) defined by a top split--gate,\cite{Nakaharai2011} where the filling factor under the top--gate depends both on the back--gate and top--gate voltage, in our geometry with the split--gate buried in PMMA under the graphene ribbon, the filling factor over the split--gate $\nu_{SG}$ is independent of the back--gate voltage. The split--gate screens the potential of the back--gate, hence the lines separating subsequent filling factors are vertical. In Fig.~\ref{scan}(a) the horizontal dashed lines indicate the threshold for subsequent filling factors $\nu_{BG}$ in the bulk of the device, and the vertical ones, the filling factors $\nu_{SG}$ over the split--gate.

Whenever $\nu_{SG}$ and $\nu_{BG}$ are equal, the resistance is zero: charge transport through the device occurs without backscattering. Indeed, Fig.~\ref{scan}(a) shows four rectangular regions that show value zero and align in the diagonal, corresponding to $\nu_{BG} = \nu_{SG}$. 

The values in the other regions are between 0 and 1 whenever these two filling factors take different values. This phenomenology was also observed in local top--gate\cite{Williams2007,Huard2007,Ozyilmaz2007} and top split--gate\cite{Nakaharai2011} devices. The resistance values are given by\cite{Ki,Nakaharai2011}
\begin{equation}
R_{xx}= \frac{h}{e^2} \frac{ \left| \nu_{BG}-\nu_{SG} \right| }{\left| \nu_{BG} \right| \left| \nu_{SG} \right| }.
\label{eq:npn}
\end{equation}

Note, however, that in the experimental results of Fig.~\ref{scan}, some of the rectangles are divided into two regions with different resistance values. We can explain this by noticing that the filling factor $\nu_{QPC}$ in the middle of the 400~nm--wide QPC may differ from $\nu_{SG}$. For the case $\nu_{QPC}=\nu_{BG}$, the QPC is open for all edge states coming from the source, and the resistance is zero. When $\nu_{QPC}=\nu_{SG}$, the device behaves like a unipolar or bipolar junction, and the resistance takes the values given by Eq.~(\ref{eq:npn}). 

In the most general case, when all three filling factors have a different value, the resistance deviates from the value given by Eq.~(\ref{eq:npn}). We can see in Fig.~\ref{scan}(a) that in many rectangles there are two different plateaus whose values will be calculated in the following section. The red dashed lines in Fig.~\ref{scan} indicate the transition between subsequent $\nu_{QPC}$: these lines are inclined, since the filling factor in the constriction opening depends on {\em both} back--gate and split--gate voltages. One can see that for a fixed set of the three filling factors the resistance is more or less constant. At the transition between subsequent $\nu_{BG}$, it rises because the Fermi energy is close to the Landau level, and instead of perfect edge states propagating in the device, we have a significant backscattering, giving rise to a higher resistance. 

The possibility to achieve a $\nu_{QPC}\neq\nu_{BG}$ is further supported by the fact that the above-cited deviations are only observed for narrow split--gates while a different behavior occurs for instance in the case of the 800~nm--wide split--gate devices (see Supplementary Information). Indeed, for wide--gap split--gates, fringe field of the finger electrodes is not expected to be sufficiently strong to induce a filling factor different from $\nu_{BG}$ in the middle of the constriction. This heuristic assumption will be better justified by the numerical simulations reported in section~\ref{SP-model}.

\section{\label{discussion} Discussion}

\begin{figure}[tb!]
  \begin{center}
   \includegraphics[width=0.6\columnwidth]{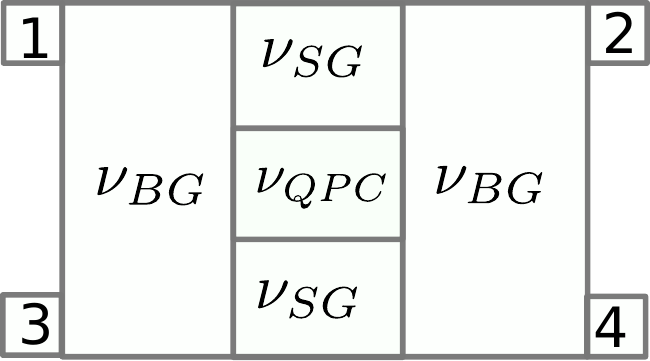}
  \end{center}
  \caption{\label{geom_} Sketch of the device geometry adopted in the Landauer--B\"uttiker model. Four contacts and regions with three distinct filling factors $\nu_{BG}$, $\nu_{SG}$, and $\nu_{QPC}$ are assumed.}
\end{figure}

The experimental resistance pattern of Fig.~\ref{scan}(a), as well as the results of the Schr\"odinger-Poisson simulation reported in the next section, can be explained in a simple model based on current conservation and edge-mode equilibration. In the model we consider a graphene device including a back--gate and a split--gate which can induce regions with various filling factors: $\nu_{BG}$ in the bulk, $\nu_{SG}$ above the split--gate electrodes, and $\nu_{QPC}$ in the QPC opening (see Fig.~\ref{geom_}). In the experiment, longitudinal resistances are obtained in a four-wire scheme by measuring the longitudinal voltage drop $V_{xx}$ in the presence a current bias $I_{SD}$ between the source to drain contacts. Four leads connected to the graphene devices are thus included in the model in order to reproduce the experimental results. The resulting device geometry, including the contacts and the various local filling factors, is show in Fig.~\ref{geom_}.

For the calculation of resistances using the Landauer--B\"uttiker formalism,  we calculate the values of conductance $G_{pq}$ for the electron flow
from terminal $q$ to $p$. Next we  build a $\boldsymbol{G}$--matrix~\cite{datta}
\begin{widetext}
\begin{equation}
\boldsymbol{G}= \left(
  \begin{array}{c c c}
	G_{12}+G_{13}+G_{14} 	&	-G_{12}		& 	-G_{13}		\\
	-G_{21} 		&  G_{21}+G_{23}+G_{24}	& 	-G_{23}		\\
	-G_{31}			&	-G_{32}		& 	G_{31}+G_{32}+G_{34}	\\
  \end{array}
 \right)
\end{equation}
\end{widetext}
and  invert it to obtain the matrix $\boldsymbol{R} = \boldsymbol{G}^{-1}$. The longitudinal resistance $R_{12,34}$ is obtained with the current flowing from lead 1 into 2 with the voltage drop measured 
between leads 3 and 4, and equals $R_{31}-R_{32}$\cite{datta}, where $R_{ij}$ is the ${\bf G}^{-1}$ matrix element of  $i$-th row and $j$-th column. Due to the symmetry in our system $G_{24} = G_{31}$, and $G_{34} = G_{21}$. For $B>0$ and $\nu_{BG}>0$, all carriers from lead 3 flow into lead 1 with the Lorentz force keeping the current at the left edge of the sample with respect to the direction of the charge flow: 
$G_{13} = \nu_{BG} e^2/h$, $G_{23} = 0$. We also assume that all carriers from lead 2 flow into lead 4, which gives $G_{32} = 0$, $G_{12} = 0$. Finally, $G_{14} = 0$, as the carriers from lead 4 can never reach lead 1.

\begin{figure}[tb!]
   \includegraphics[width=\columnwidth]{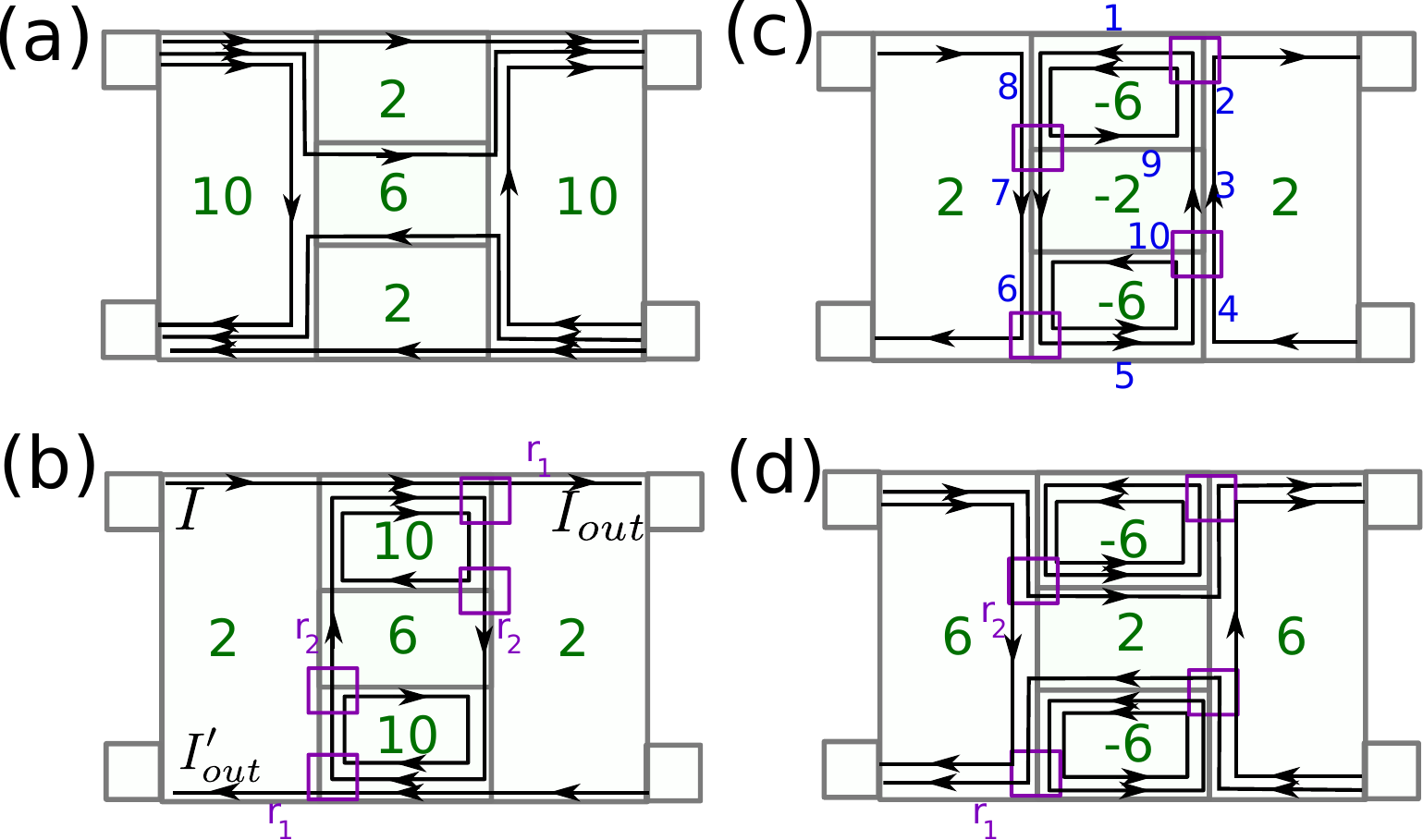}
   \caption{\label{geom} Edge connectivities as a function of the three filling factors $\nu_{BG}$, $\nu_{SG}$, and $\nu_{QPC}$. Each sketch implies a different edge propagation, splitting, and equilibration configuration and thus a different analytical formula for the longitudinal resistivity. Configurations with opposite chiralities are not reported explicitily. Black arrows indicate the currents for each of the spin-degenerate available transport modes and their chiarality. In order to provide a specific example, the green numbers indicate one of the possible filling factor combinations giving rise to the sketched connectivity scheme. In panels (b), (c), and (d), edge equilibration in co-propagating modes plays a crucial role: the purple squares indicate the points where current branches; in the model current is assumed to partition equally on all available channels, i.e. edge modes are assumed to completely equilibrate before the branching.}
\end{figure}

Depending on the relative chirality of the bulk, split--gate, and QPC regions, there are four regimes to be considered, as shown in Fig.~\ref{geom}.
For the calculation of the conductance matrix elements, we proceed in a similar way as in Ref.~\onlinecite{Ozyilmaz2007} on $n$--$p$--$n$ junctions, 
which however covered a simpler case with the edges only between pairs of regions of varied $\nu$. The numerical modeling (see the next section)
indicates that here mode mixing needs to cover all border lines between the three separate filling-factor regions.
In Ref.~\onlinecite{Zimmermann} for a top split--gate the anomalous Hall plateaus could be explained by the equilibration 
of the $N=0$ Landau level. The results for the present sample call for mixing of all the subbands.

In the first case (Fig.~\ref{geom}(a)) transmission of $\nu_{QPC}$ edge modes of the incident $\nu_{BG}$ modes occurs. Therefore, the resistance is the same as in a bipolar junction with filling 
factor $\nu_{QPC}$ in the middle area, and it equals
\begin{equation}
R_{12,34}= \frac{h}{e^2} \frac{ \left| \nu_{BG} \right|- \left| \nu_{QPC}\right| }{ \left| \nu_{BG}\right| \left|\nu_{QPC} \right| } .
 \label{eq:unipolar1}
\end{equation}
For the special case shown in Fig.~\ref{geom}(a), Eq.~(\ref{eq:unipolar1}) gives $R_{12,34}= 1/15$.

In the second case (Fig.~\ref{geom}(b)), edge mode mixing occurs, and the current in the spots marked by purple 
rectangles is partitioned equally between available modes. 
The probabilities of the current going into one of the directions shown in Fig.~\ref{geom}(b) are $r_1 = \frac{|\nu_{BG}|}{|\nu_{SG}|}$ and $r_2 = \frac{|\nu_{QPC}|-|\nu_{BG}|}{|\nu_{SG}|-|\nu_{BG}|} $.
We label the current at each side of the split--gate by 1, 2, ...,  10 as shown in Fig.~\ref{geom}(c).
The currents satisfy the following relations: $I_2=(1-r_1)I_1$, $I_3=r_2 I_2$,  $I_9=(1-r_2) I_2$, 
$I_6=(1-r_1)I_5$, $I_7=r_2 I_6$, and $I_{10}=(1-r_2) I_6$. 
Assuming that a current $I$ flows in the 1st lead, we find the following current conservation
rules:
$I+I_8=I_1$, $I_7+I_9=I_8$, $I_3+I_{10}=I_4$, and $I_4=I_5$. We construct a system of 10 equations 
and find the outflowing currents $I_{out}=I_1-I_2$ and $I'_{out}=I_5-I_6$. 
Then, the conductances are: $G_{21}=\frac{2e^2}{h}|\nu_{BG}| I_{out}/I$ and $G_{31}=\frac{2e^2}{h}|\nu_{BG}| I'_{out}$. 
Inserting this information into the $\boldsymbol{G}$ matrix and inverting it, we find:
\begin{equation}
R_{12,34}= \frac{h}{e^2} \frac{ \left( \left|\nu_{QPC}\right|-\left| \nu_{BG}\right| \right) }{\left| \nu_{BG}\right| \left|\nu_{QPC}\right| },
\label{eq:unipolar2}
\end{equation}
which is equivalent to a unipolar junction with a filling factor satisfying $|\nu_{QPC}|>|\nu_{BG}|$ in the middle area. For the special case shown in Fig.~\ref{geom}(b), Eq.~(\ref{eq:unipolar2}) yields $\frac{1}{3}$.

The remaining two cases are shown in Figs.~\ref{geom}(c) and (d).
The resistances are (for the calculation see Supplementary Information):
\begin{equation}
R_{12,34}= \frac{h}{e^2} \frac{\left( \left| \nu_{SG}\right|+ \left|\nu_{BG}\right|\right) \left(\left| \nu_{QPC}\right|+\left| \nu_{BG}\right|\right) }{\left| \nu_{BG}\right| \left( \left( \left| \nu_{SG}\right|-\left| \nu_{BG}\right| \right) \left| \nu_{QPC}\right| +2\left| \nu_{BG}\right| \left| \nu_{SG}\right| \right) }
\label{eq:bipolar1}
\end{equation}
for the case presented in Fig.~\ref{geom}(c), and 
\begin{equation}
R_{12,34}= \frac{h}{e^2} \frac{\left( \left| \nu_{SG}\right| +\left| \nu_{BG}\right|\right) \left(\left| \nu_{QPC}\right|-\left|\nu_{BG}\right|\right) }{ \left|\nu_{BG}\right| \left( \left(\left| \nu_{SG}\right|-\left|\nu_{BG}\right|\right) \left|\nu_{QPC}\right|-2\left|\nu_{BG}\right|\left|\nu_{SG}\right| \right) }
\label{eq:bipolar2}
\end{equation}
for the case shown in Fig.~\ref{geom}(d).

Summarizing, the resistance is given by Eqs.~(\ref{eq:unipolar1})--(\ref{eq:bipolar2}) for the four configurations of the filling factors shown in Fig.~\ref{geom}. 
Note that always $|\nu_{QPC}|\le|\nu_{SG}|$. Fig.~\ref{scan}(b) shows a schematic plot with the calculated filling factors in each part of the device. There is a good agreement observed with the experimental data of Fig.~\ref{scan}(a).

\section{\label{SP-model} Schr\"odinger-Poisson model}

The previous section described a simple model based on the Landauer-B\"uttiker formalism that can explain the observed resistance pattern, provided one makes the assumption of three independent filling factors. In this section, we shall show that such an assumption can indeed be directly derived from a Schr\"odinger-Poisson model of the device.\cite{Lis,Bednarek} Based on the resulting potential landscape, the quantum transport problem is solved numerically. The calculation is performed using a wavefunction matching method and yields the actual current paths and $G_{pq}$ values from the scattering wave functions at the Fermi level. These parameters are then used to calculate the longitudinal resistance.

\subsection{Description of the model}

The numerical calculation is performed on a graphene region with zigzag horizontal and armchair vertical edges. The constriction region has a width of $197.2$~nm and a length of $443.2$~nm. In the simulation, the leads are $24.6$~nm long and $17.4$~nm wide, which corresponds to a width of 40 atoms across the ribbon. This gives a total length of the device with contacts of 492.4~nm. For the simulation we use the scaling approach of Ref.~\onlinecite{Rickhaus}, with a scaling factor $s_f$. The qualitative results of the model do not change when $s_f$ is increased beyond $4$ and only depend on the configuration of the filling factor regions. Simulations were thus performed with $s_f=4$.
The simulated system is still smaller than the real one: this is compensated by performing calculation with (i) smaller voltages in order to induce the same electric field values as in the larger device (see discussion below)
and (ii) larger values of the external magnetic field that drive the same number of available edge channels.

\begin{figure}[tb!]
  \begin{center}
   \includegraphics[width=0.7\columnwidth]{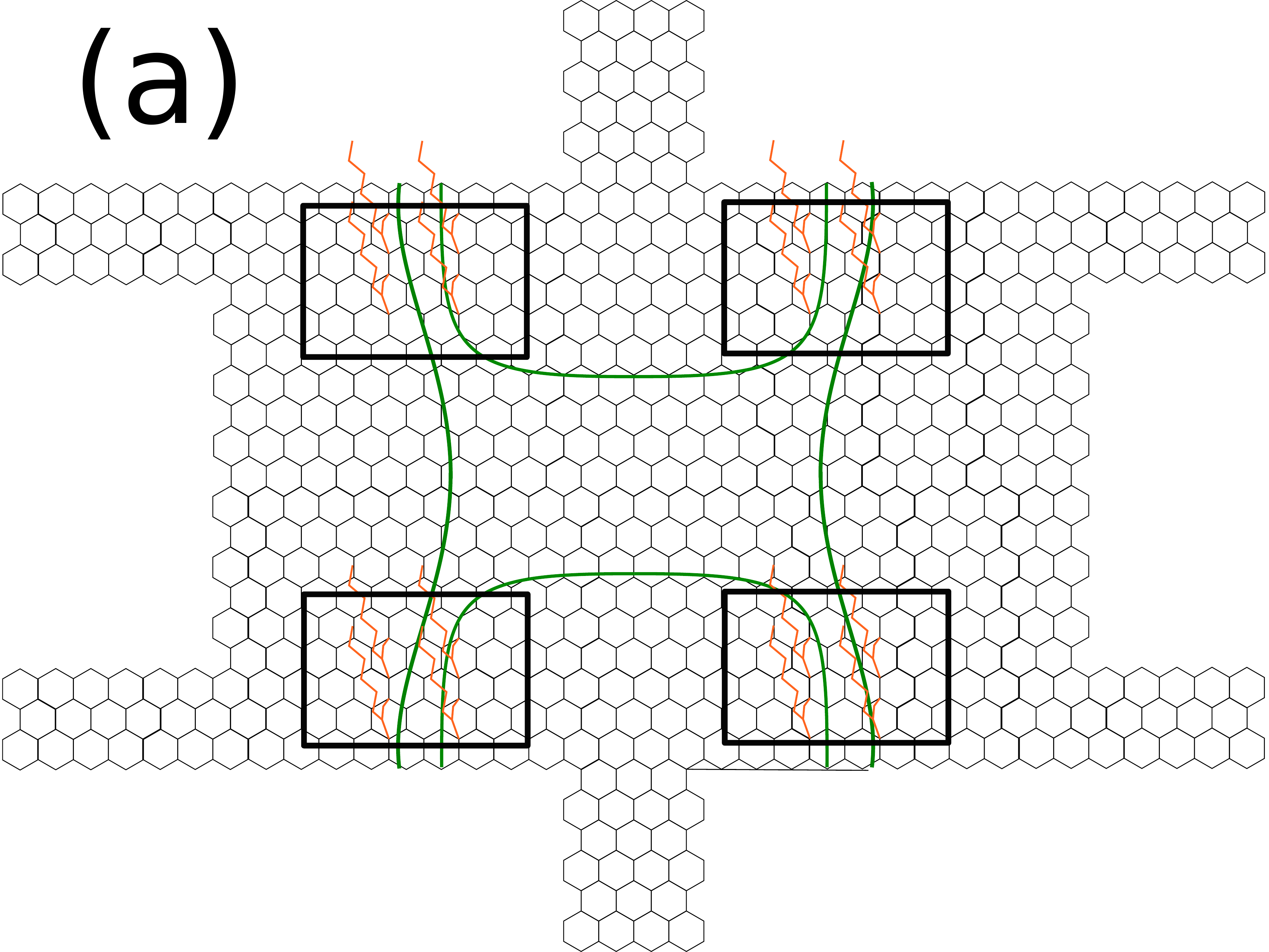}
   \includegraphics[width=0.8\columnwidth]{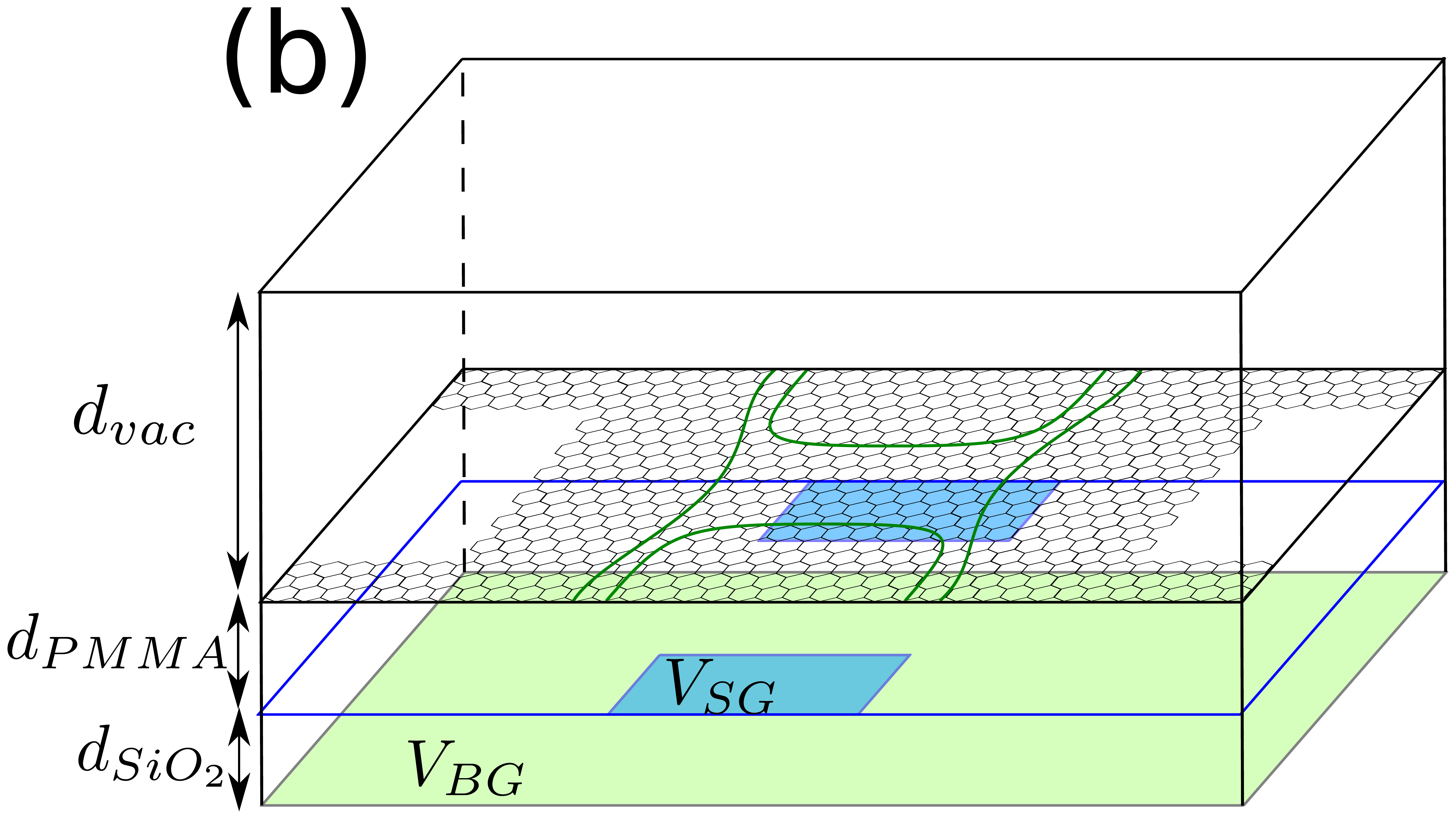}
  \end{center}
  \caption{\label{lat} Schematic drawing of the studied system. (a) Scheme of the nanoribbon with four horizontal zigzag leads, two vertical armchair B\"uttiker probes, and four B\"uttiker probes connected to the interior of the nanorribon near the $n$--$p$ junctions, highlighted in orange. Green lines show schematic isolines of the potential. (b) The computational box for the Laplace problem. Voltage $V_{BG}$ is applied to the back--gate (bottom of the computational box), and $V_{SG}$ to the split--gate coloured in blue. The dimensions used for the simulation are: $d_{SiO_2} = 27.6$~nm, $d_{PMMA} = 9.8$~nm, $d_{vac} = 208.6$~nm.}
\end{figure}

In order to evaluate the electrostatic potential energy experienced by the electron gas within the graphene device, we solved the Poisson equation for the considered geometry with the electron density given by the Thomas--Fermi approximation,\cite{Brey} and found that the influence of the space charge for the profile of the potential landscape is negligible. The potential is thus determined by the Laplace equation for the system of electrodes given in Fig.~\ref{lat}. We consider a computational box of $492.4 \times 197.2 \times 246$~nm$^3$ in the $x\times y\times z$ direction, respectively, where the size in $x$ includes the scattering region and the left and right leads. The back--gate and the split--gate are placed at $z=0$ and $z=27.6$~nm, respectively, and the nanoribbon at $z=37.4$~nm. The split--gate (size of each gate finger $165.2$~nm in $x$ and $68.1$~nm in $y$) is placed in the middle of the ribbon's length, with a spacing between the gate fingers of $61$~nm. At each electrostatic gate we use Dirichlet boundary conditions for applied potentials and at the side and top walls of the computational box Neumann boundary conditions with zero electric field, which at the top side of the box is justified by the charge neutrality of the system, and at the lateral sides by the symmetry of the system far from the split--gate. The value of $d_{vac}$ is chosen so that the potential within the graphene layer does not change any further upon increasing the box height.

For the calculations, we use the tight--binding Hamiltonian
\begin{equation}
   H=\sum_{\langle i,j\rangle }t_{ij} \left(c_i^\dagger c_j+c_j^\dagger c_i\right)+\sum_i V\left({\bf r}_i\right) c_i^\dagger c_i, 
\label{eq:dh}
\end{equation}
where $V({\bf r}_i)$ is the external potential at $\mathbf{r}_i$, the position of the $i$th atom, and the first summation runs over the nearest neighbors. The magnetic field is taken into account by Peierl's  substitution in the hopping parameter, 
\begin{equation}
t_{ij} = t \exp \left( \frac{2\pi i}{\phi_0} \right) \int_{\mathbf{r}_i}^{\mathbf{r}_j} \mathbf{A}\cdot\mathbf{dl}, 
\end{equation}
where $t$ is the hopping parameter and $\phi_0=\frac{h}{e}$ the flux  quantum. For a magnetic field perpendicular to the graphene plane $\mathbf{B}=(0,0,B)$, we use a Landau gauge $\mathbf{A}=(-yB,0,0)$. As already mentioned, a scaling approach is used,\cite{Rickhaus} with scaling  condition $a=a_0 s_f$ and $t=t_0/s_f$, where the scaling factor is $s_f=4$, $t_0=-2.7$~eV is the unscaled hopping parameter, while $a_0=2.46$~{\AA } is the graphene lattice constant. The rescaled magnetic field is $B=B_0 s_f^2$, with $B_0$ being the magnetic field characterizing the real sample. The ratio of the ribbon width $l$ to magnetic length $l_B$ equals $l/l_B= 56.76$, with magnetic length $l_B=\sqrt{\frac{\hbar}{eB_0}} = 26$~nm$/\sqrt{B_0[\mathrm{T}]}$. Zero temperature is assumed. We determine the filling factor as the number of Landau levels below the Fermi energy.

To solve the scattering problem, we use wave function matching (WFM). The details of the computational method are described in Ref.~\onlinecite{Kolasinski2016}. The transmission probability from terminal $l$ to mode $m$ in terminal $k$ is
\begin{equation}
T_{kl}^m = \sum_{ n } \left| t_{mn}^{kl}\right|^2,
\label{eq:transprob}
\end{equation}
with $t_{mn}^{kl}$ being the probability amplitude for the transmission from the mode $n$ in terminal $l$ to mode $m$ in the terminal $k$.

Fig.~\ref{geom_} shows the labeling of the leads in the model system.
To compute $R_{12,34}$, we construct a conductance matrix $\mathbf{G}$ of dimension $N-1$, where $N$ is the number of terminals, and calculate it as a sum over the modes:
\begin{equation}
G_{kl} = G_0 \sum_{ m } T_{kl}^m,
\end{equation}
where $G_0=\frac{2e^2}{h}$ is the conductance quantum. Then, we use the following formula to relate the current $I_k$ in terminal $k$ to the voltages in all terminals:
\begin{equation}
I_k = \sum_{ l } G_{kl} \left( V_k-V_l \right),
\end{equation}
where $V_k$ and $V_l$ are voltages in terminals $k$ and $l$, respectively. 
The resistance calculation proceeds as explained in section IV.

In view of a comparison between numerical results and actual experimental data, it is crucial to comment on the magnitudes of the gate voltages in relation with the scaling method adopted in the model. Indeed transport properties are simulated in a {\em scaled} atomistic model. As a consequence, gates were assumed to be closer to the graphene plane than in the actual devices, which reduces the number of mesh elements in the finite difference solver. In particular, in the simulation the dimensions of the system are approximately 10 times smaller than in the experiment. The gate--voltage--to--energy conversion factor is inversely proportional to the distance between the electron confinement area and the gates. Hence, because of the reduced distance between the gates and the graphene, the gate voltages inducing a specific filling factor distribution are accordingly smaller.

\subsection{Numerical results}

\begin{figure*}[tb!]
   \includegraphics[width=\textwidth]{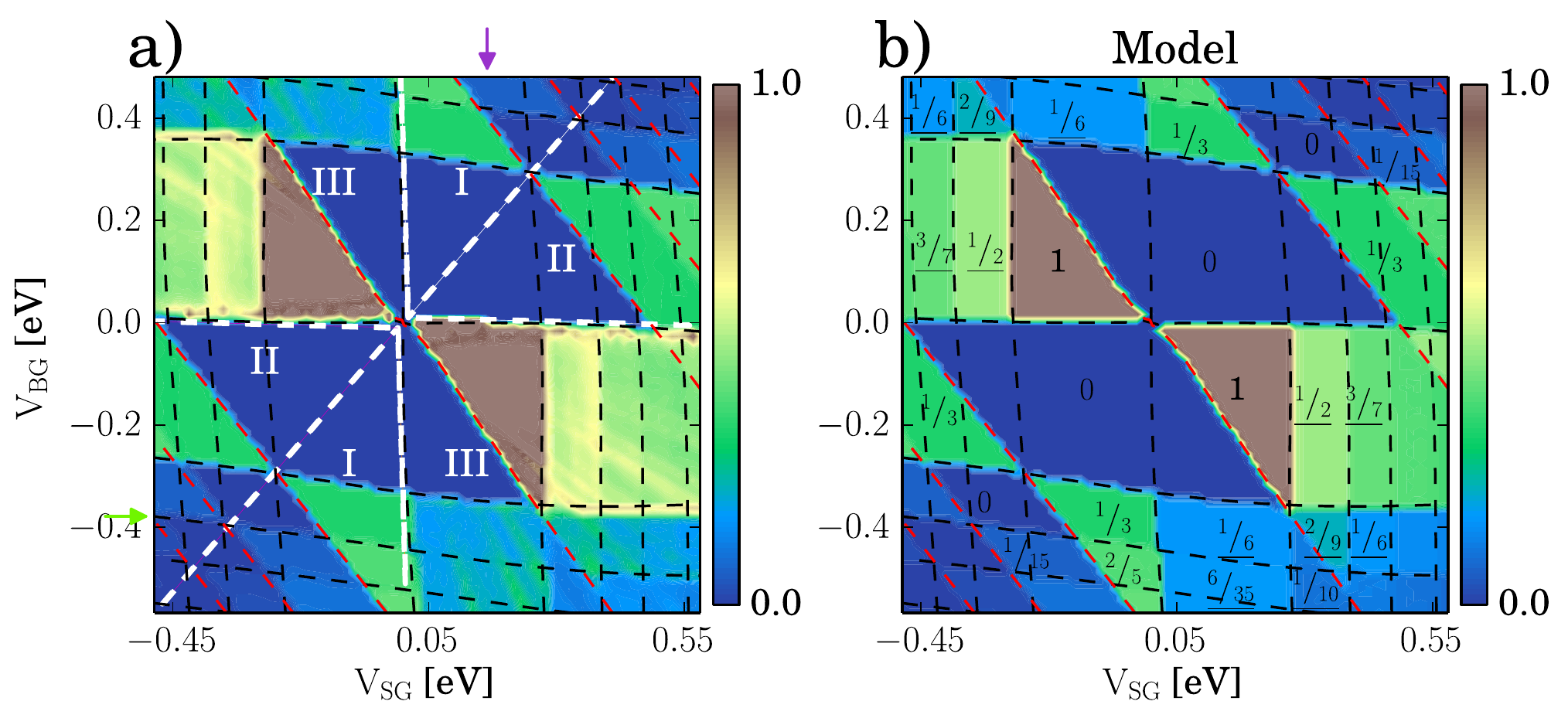}
   \caption{\label{scanSym} (a) Simulated longitudinal resistance (in units of $h/e^2$) as a function of $V_{BG}$ and $V_{SG}$. (b) Resistance values calculated according to Eqs.~(\ref{eq:unipolar1})--(\ref{eq:bipolar2}) (in units of $h/e^2$). The non-linear shape of the boundaries of regions at fixed $\nu_{BG}$ is caused by the limited size of the computational box which cause the potential in the leads to slightly depend on $V_{SG}$.}
\end{figure*}

When transport is fully coherent, no equilibration occurs between co-propagating edge modes and the model cannot correctly reproduce the observed behavior (see Supplementary Information). In order to induce partitioning of the current, dephasing B\"uttiker virtual probes were introduced, as visible in Fig.~\ref{lat}(a). These are used as voltage probes, i.e. a zero net current flow is assumed at each of the probes. Electrons entering the probe equilibrate in the reservoir, and emerge from it with a different phase. The choice of the positions of such artificial probes is crucial for edge-mode mixing. According to the theoretical models,\cite{Ozyilmaz2007, Williams2007, Abanin2007} equilibration in the bipolar junction takes place along the $n$--$p$ interface, and in the case of a unipolar junction, shown in 
Fig.~\ref{geom}(b), along the edge in the central, split--gate region. For the latter case, the addition of voltage probes connected to the region over the split--gate results in resistance values that are in very good quantitative agreement with the values given by Eq.~(\ref{eq:unipolar2}). For the former case, no configuration of the probes in the plane of the graphene nanoribbon gives values close to what is predicted by Eqs.~(\ref{eq:bipolar1}) or (\ref{eq:bipolar2}). However, we obtained good results for the probes connected to the interior of the nanoribbon
near the $n$--$p$ junctions,\cite{Metalidis,chen} as shown in Fig.~\ref{lat}(b). The position and length of the probes is set so that most of the currents flowing along the junction can reach the probes. 

The final position and size of the probes is discussed in the following. Two probes are armchair nanoribbons of 35 atoms width (i.e.~they approximately have the same width as the leads) connected to the region over the split--gate, and are assumed to be semi--infinite in the $y$ direction. Further four probes are semi--infinite in the $z$ direction and attached to the ribbon in the vicinity of the split--gate, marked in orange in Fig.~\ref{lat}(a) and consisting of $60$ zigzag chains attached to the graphene plane 
within four areas marked schematically by black rectangles in Fig.~\ref{lat}(a) (see Supplementary Material for further details).

We calculate $R_{12,34}$ as a function of potentials $V_{BG}$ and $V_{SG}$ at the back--gate and split--gate, respectively (see Fig.~\ref{lat}). The calculated resistance is presented in Fig.~\ref{scanSym}(a). The horizontal dashed lines indicate the transition between subsequent filling factors $\nu_{BG}$ in the bulk of the nanoribbon, the vertical ones between filling factors $\nu_{SG}$ in the split--gate, and the red dashed lines between filling factors $\nu_{QPC}$ in the middle of the QPC. Following the nomenclature of Ref.~\onlinecite{Ozyilmaz2007}, the plot is divided into sectors, depending on the relationship between the filling factors. In the sectors between dashed white lines labeled with I, where the filling factors have the same sign $\nu_{BG}\cdot\nu_{SG}>0$ and $\nu_{BG}\ge\nu_{SG}$, the device is in the edge state transmission regime. In the sectors labeled II, partial equilibration occurs. As shown in section~\ref{discussion}, in both aforementioned regimes, the device behaves like a unipolar junction with the resistance governed by $\nu_{QPC}$. In the most unique case, in sectors III, it is governed by the full equilibration process. Whereas in sectors I and II the resistance is independent of $\nu_{SG}$, in sector III it depends on all three filling factors. This is nicely consistent with what was observed in the experimental data (see Fig.~\ref{scan}(a)).

\begin{figure*}[tb!]
  \begin{center}
   \includegraphics[width=0.48\textwidth]{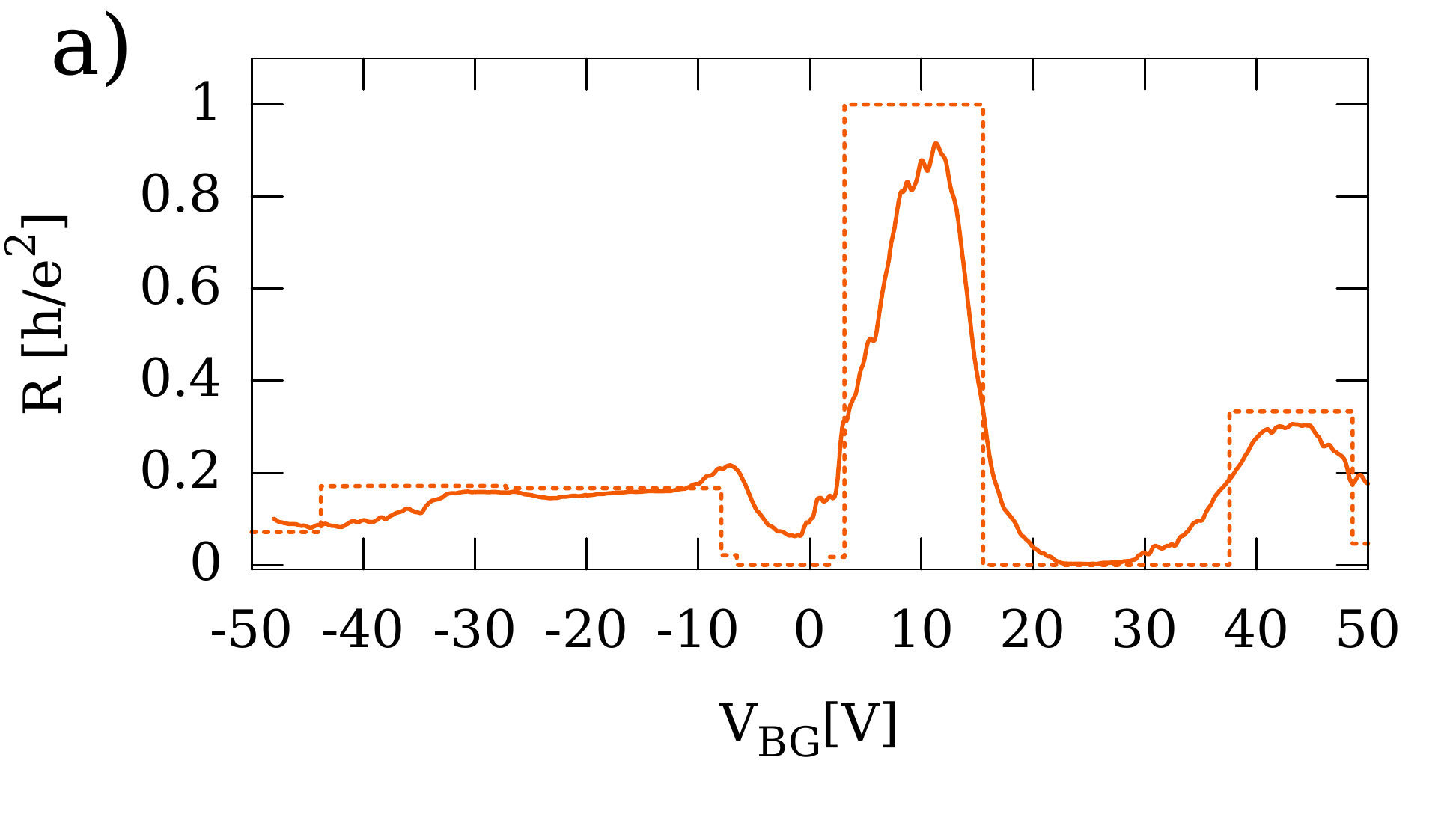}
   \includegraphics[width=0.48\textwidth]{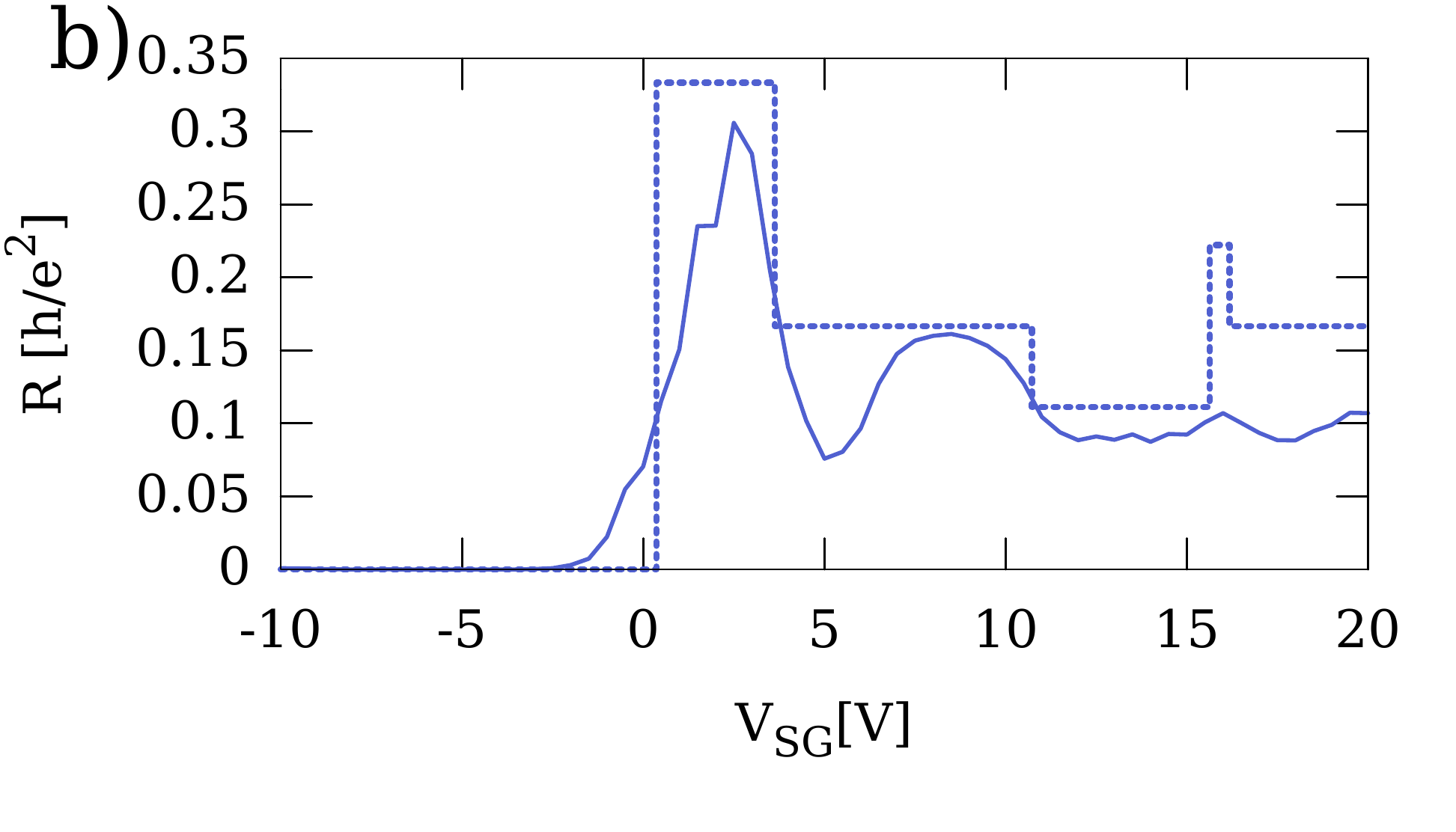}
   \includegraphics[width=0.48\textwidth]{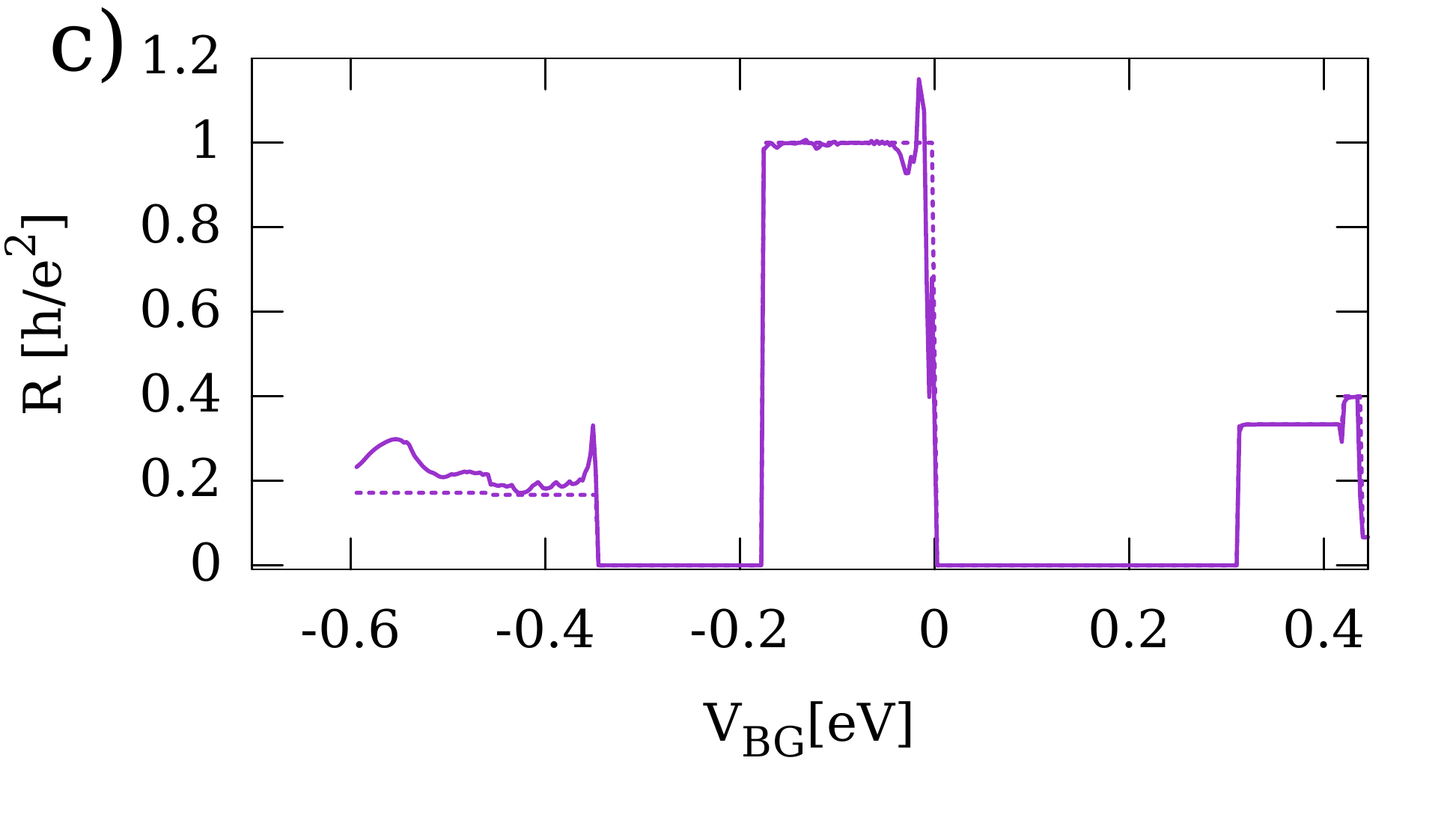}
   \includegraphics[width=0.48\textwidth]{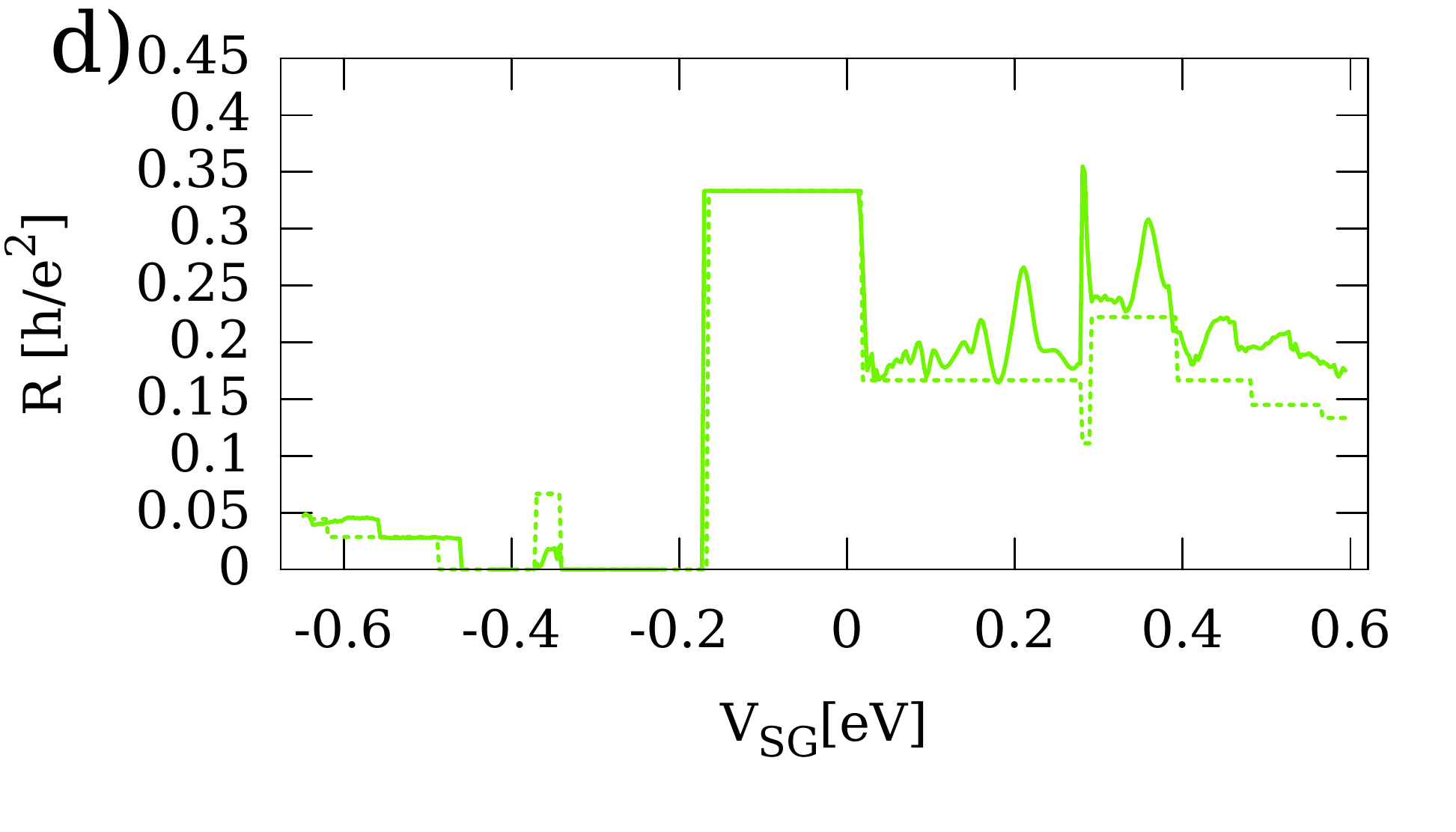}
  \end{center}
  \caption{\label{cross} The cross--sections of the experimental (a,b), and the simulated (c,d) data, indicated by arrows in Figs.~\ref{scan} and \ref{scanSym}. The color of the arrow corresponds to the line color. Dashed lines show the model values of resistance given by Eqs.~(\ref{eq:unipolar1})--(\ref{eq:bipolar2}). (a) $\nu_{SG} = 2, V_{SG} = 8$~V, (b) $\nu_{BG} = -6, V_{BG} = -13$~V, (c) $\nu_{SG} = 2, V_{SG} = 0.135$~eV, (d) $\nu_{BG} = -6, V_{BG} = -0.39$~eV.}
\end{figure*}

The corresponding model values obtained with the Landauer--B\"uttiker formalism discussed in section~\ref{discussion} are shown in the 2D plot in Fig.~\ref{scanSym}(b). 
The underlined numbers are the values for $\nu_{BG}\ne\nu_{SG}\ne\nu_{QPC}$ in the full equilibration regime. There is a perfect agreement between the two plots in Figs.~\ref{scanSym}(a) and \ref{scanSym}(b) for region I, which would be the case even without any B\"uttiker probes. In the edge state transmission regime, no equilibration occurs, and only $\nu_{QPC}$ modes out of incoming $\nu_{BG}$ modes can pass the split--gate, therefore no phase--randomizing is needed to obtain the expected resistance. 
In regime II the resistance also coincides with the model values, as the high magnetic field used for the simulation forces all the incoming electrons to the side--probes, where they equilibrate. In sector III, where equilibration occurs along the BG--SG interface, the agreement with the model data is good for $\nu_{QPC}=\nu_{SG}$, but the simulated values for $\nu_{QPC}\ne\nu_{SG}$ exceed slightly the model ones. The reason is that the magnetic field perpendicular to the plane of the ribbon does not push the electrons into the probes in $z$, so that not all electrons enter the probes and a small fraction of electrons does not equilibrate. For $\nu_{QPC}=\nu_{SG}$, for which the edge--state partitioning takes place at the point where the junction interface meets the edge, this does not yield a large deviation from the model resistance values. However, in the more complex case of $\nu_{QPC}\ne\nu_{SG}$, the random partitioning is expected to take place also between the SG and QPC regions, the effectiveness of which is somewhat smaller in the simulation.

In order to underline the excellent agreement between experiment, model, and simulation more thoroughly, we plot cross--sections of the longitudinal resistance. In Figs.~\ref{cross}(a) and (b) cross--sections of the experimental data are plotted with orange and blue lines for the $V_{SG}$ or $V_{BG}$ values indicated by arrows of the same color in Fig.~\ref{scan}(a). The dashed line shows the value of resistance resulting from the model. The resistance reaches plateaus or at least gets close to the expected value for almost every set of $\nu_{BG},\nu_{SG},\nu_{QPC}$ in the cross--sections. A particularly good agreement for $V_{SG}=8$~V is seen for the plateaus $\frac{1}{10}$, $\frac{6}{35}$, $\frac{1}{6}$, i.e.~with $\nu_{SG}\ne\nu_{BG}\ne\nu_{QPC}$. For the cross--section at $V_{BG}=-13$~V, the resistance gets close to the model values $\frac{1}{3}$, $\frac{1}{6}$, and $\frac{1}{9}$, but it does not reach the plateau $\frac{2}{9}$ for $\nu_{SG}=6$, $\nu_{BG}=-2$, and $\nu_{QPC}=2$, and the plateau $\frac{1}{6}$ for $\nu_{SG}=10$, $\nu_{BG}=-6$, and $\nu_{QPC}=2$. A too small resistance is obtained for almost every case in the partial equilibration regime.

In Figs.~\ref{cross}(c) and (d),  cross--sections of the simulated data are presented for $V_{SG}$ indicated by arrows in Fig.~\ref{scanSym}(a) with the same color as the plot line. The agreement with the model is very good, and only in the case of three different filling factors in the full--equilibration regime, there is a small discrepancy. However, even in this case the curve follows the model line. 
The shape of the simulated and measured curves clearly is similar, one difference being the energies at which subsequent bands enter the transport, and the slope of the red dashed contours in Fig.~\ref{scan}, which gives rise to the occurrence of different sets of the three filling factors.

\section{Summary and conclusions}

We have analyzed the behavior of buried split--gate graphene devices which are suitable for scanning-probe-microscopy experiments. Differently from studied experimental configurations,~\cite{Nakaharai2011} the measurements of the longitudinal resistance taken at $B=0$ indicated a screening of the  back--gate by the split--gate within the 
QPC areas: this implies that no crosstalk between the two is observed. Moreover, the measurements taken in the QH regime as well as the numerical simulation yield a resistance pattern indicating the clear presence of regions with three distinct filling factors. The resistance pattern can be explained with the mode equilibration of the edge currents that involves all three regions of varied filling factors and all modes participating in the current flow.

\begin{acknowledgments}
We acknowledge funding from the Italian Ministry of Foreign
Affairs, Direzione Generale per la Promozione del Sistema
Paese, and from the Polish Ministry of Science and Higher Education, Department of International Cooperation (agreement on scientific collaboration between Italy and Poland).
Financial support
from the CNR in the framework of the agreements on scientific
collaborations between CNR and CNRS
(France), NRF (Korea), and RFBR (Russia) is acknowledged.
Funding from the European Union
Seventh Framework Programme under Grant Agreement No.
696656 Graphene Core1 is acknowledged. 
The theoretical side of this work was supported by the National Science Centre
(NCN Poland)  according  to  decision  DEC-2015/17/B/ST3/01161,
and by PL-Grid Infrastructure. A. M.-K.~acknowledges the doctoral stipend Etiuda funded by the National
Science Centre (NCN) according to DEC-2015/16/T/ST3/00264. S.~G.~acknowledges support by Fondazione Silvio Tronchetti Provera.
\end{acknowledgments}


%

\end{document}